\documentclass[aps,prl,showpacs,twocolumn]{revtex4-1}
\usepackage{enumitem}
\usepackage{amsfonts}
\usepackage{ulem}
\usepackage[T1]{fontenc}
\usepackage[utf8]{inputenc}
\usepackage{hyperref} 
\usepackage{epsfig}
\usepackage{color}
\usepackage{multirow}
\usepackage{float} 
\usepackage{amsthm,graphics,amssymb,amsmath,latexsym}

\usepackage{fancyhdr}
\usepackage{dcolumn}
\usepackage{bm}
\usepackage[mathlines]{lineno}
\usepackage{mathptmx}

\begin{document}
\title{General method to perform Microcanonical Monte Carlo Simulations }
\author{G. Palma}
\affiliation{Departamento de F\'isica, Universidad de Santiago de Chile, Avda. Ecuador 3493, 917-0124 Santiago, Chile}
\author{A. Riveros}
\email{alejandro.riveros@usach.cl}
\affiliation{Departamento de F\'isica, Universidad de Santiago de Chile, Avda. Ecuador 3493, 917-0124 Santiago, Chile}
\date{\today}
\begin{abstract}

Monte Carlo simulations have boosted the numerical study of several different physical systems and in particular, the canonical ensemble has been especially useful because of the existence of easy and efficient relaxation algorithms required to  minimize the energy, the relevant extensive thermodynamic variable appearing in the probability distribution, which drives the system after a thermalization process to equilibrium. Nevertheless, the nature does not know about statistical ensembles and therefore it is desirable and a theoretical challenge to show how to perform efficient numerical simulations in the microcanonical ensemble. In this article, we present a method based on the concepts of configurational temperature estimator \cite{Rugh,GDP} and on stochastic dynamics to do it. The method is independent of the Monte Carlo update strategy, and can be implemented for both local update or cluster algorithms. We illustrate the main features of the method by performing a numerical simulation of the planar interacting classical spin system, known as the two-dimensional XY-model. 
\end{abstract}


\maketitle

In the last decades the study based on numerical simulations of condensed matter models and lattice gauge theories have made an important progress in part because of the increasing capabilities of computers and also due to the development of efficient algorithms, including the cluster and renormalization group based algorithms \cite{S,PZ}. Among the numerical strategies, most of the Monte Carlo (MC) simulations come from the seminal article of Metropolis et al \cite{Metropolis}. From the point of view of the Statistical Mechanics, instead of computing the dynamics of a system, which represents a very difficult problem and in the vast majority of the cases leads to an unreachable highly complexity task, the description of a physical system amounts computing thermal averages of the observables in the phase or configurational space of the system. The equivalence of both approaches is guaranteed by the quasi-ergodic theorem. On the other side, the generation of a sequence of configurations is achieved by a Markovian process which fulfils the necessary conditions of accessibility and microscopic reversibility, such that after a transient, the generated configurations are weighted according to the equilibrium probability distribution, e.g. the Gibbs distribution for the canonical ensemble (CE).

The concept of ensembles plays an essential role in the formulation of Statistical Mechanics and amounts to distinguish the macroscopic variables held fixed from their conjugated fluctuating quantities to study the physical system. The different ensembles, for instance, the microcanonical ensemble (MCE), in which the system energy, the volume and the number of particle are held fixed $ (E,V,N) $, while in the grand canonical one the temperature, the volume and the chemical potential are held fixed $ (T,V,\mu) $, are equivalent to one another in the thermodynamic limit, at least for finite range interacting systems \cite{libroRuelle}. In spite of the fact that for certain type of critical systems the description of a hysteresis curve or a critical line could be easier or clearer achieved when using a particular ensemble, the physical content must be the same in each ensemble.

The MC simulations originated in reference \cite{Metropolis} are suited in the CE $(T, V, N)$, and the fact that the Metropolis algorithm is so easy to implement for different systems had the consequence that it has been widely used. Nevertheless, the extension of the MC simulations to the MCE has been elusive probably because of the presence of a Dirac's delta in the distribution of probability of this ensemble (see discussion below), which ensures that all states with a given macroscopic energy value $ E $, have a priori equal weight. As a consequence, this feature makes any direct naive algorithm in the configurational space very inefficient, as the surface of constant energy involved in this constraint has zero volume in this space. This difficulty has lead to alternative strategies.     

In a pioneering paper \cite{MCreutz} Creutz proposed a MC simulation in the MCE of a SU(2) lattice gauge theory, in which an extra degree of freedom, called Demon carrying an energy $ E_{D} $, is introduced into the distribution probability. This strategy is similar to the use of kinetic energy in molecular dynamics simulations \cite{Plimpton}, and it aims defining approximated random walks on the surface of constant energy $ E $. As a consequence, the energy does not remain constant, as the demon is allowed to exchange energy with the system, but fluctuates above a fixed value. Another interesting method was proposed by Ray \cite{Ray} for condensed matter systems, for which a kinetic term of the form $ \Sigma_{i} p_{i}^{2}/2 m_{i} $ is introduced such that the potential energy fluctuates, with the overall system energy held fixed. In Ref.\cite{RayFrelechoz} this strategy is extended for discrete systems, and as an illustration, applied to the two-dimensional Ising model. Aside of an arbitrary definition of the temperature by an extended virial relation to spin systems, and some relations derived from it, discrepancies in the specific heat among the results obtained in the canonical and MCE computations for a 900 spins system are reported. They are of the order of 16 percent, and are attributed to ensemble dependence description of finite size systems.

In ref.\cite{Oliveira} a MCE suited for MC simulations of lattice gas models with nearest neighbour interactions was introduced, in which both the number of particles as well as the energy are held fixed. The authors use the equivalence among different ensembles to derive expressions for the temperature and chemical potential for a fixed system energy $E$. These relations allow to compute them within the MCE by performing a MC simulation with a generalization of the Kawasaki dynamics \cite{libroKawasaki}, allowing uniquely random-distance particle-hole exchanges that preserves the energy value.    
 
In this letter we propose a novel direct and general method to perform MC simulations in the MCE without introducing external artefacts, the use of ad hoc particular algorithms or model oriented relations among ensembles. It relies on the existence of configurational estimators for the temperature and energy and a locally one-to-one equation of state $ E = E(T) $. It has the useful feature of allowing the use of efficient algorithms, including cluster algorithms \cite{SwendsenWang,Wolff}, developed in the CE to study critical systems. 
 
The aim of our method is to generate a sequence of microscopic system configurations $\vec{x}$ compatible with macroscopic given values $(E, V, N)$, which are known to be distributed according to the distribution of probability of the MCE, $P_{mc}(\vec{x}) = \delta(E- \mathcal{H}(\vec{x}))/\Omega(E,N,V)$, 
%
where $ \mathcal{H}(\vec{x})$ is the Hamiltonian of the system, $\delta(E - \mathcal{H}(\vec{x}))$ is the Dirac's delta, and $\Omega(E,V,N)$ is the total number of the microscopic states compatible with the macroscopic values $(E, V, N)$. 

The main difficulty to use the above probability distribution in a MC simulation is related to the fact that the delta is a tempered distribution with support of zero measure, instead of being a wide ranged distribution as it is in the case of the CE, and therefore it is not suited to be used to sample the configurational space.  

We propose instead a method based on the existence of a locally invertible equation of state for the system and on use the two microscopic estimators for the energy and for the temperature
\begin{equation}\label{energy_beta_estim}
\hat{E} =  \mathcal{H}( \vec{x})  \,, \hspace{0.5cm}  \hat{\beta} = \vec{\nabla} \cdot \left( \frac{\vec{\nabla} \mathcal{H}( \vec{x})\,}{\parallel \vec{\nabla} \mathcal{H}( \vec{x})\, \parallel^2 } \right),
\end{equation}
where $\vec{\nabla}$ is the gradient operator in the phase space of the system \cite{Estimador}, described as follows:

\begin{enumerate}[label=(\roman*)]
\item For the given energy value $ E_{0}$  find by a physically educated guess, lower and upper temperature bounds $ T_{l} < T_{u} $ such that their corresponding energy values $ E_{l}$ and $ E_{u} $ obtained by measuring them using conventional MC simulations in the CE fulfil $ E_{l} < E_{0} <E_{u} $. As these MC simulations are only meant to successively approximate to the desired temperature $ T_{0}$, which corresponds to the desired energy $ E_{0}$, the temperature and energy estimators given by Eq.\eqref{energy_beta_estim} will be used to monitor the thermalization path of the ergodic dynamics underlying the Markov process involved.  

\item The first iterative approach to $T_{0}$ will be the value obtained by a linear interpolation between the points $(T_l,E_l)$ and $(T_u,E_u)$ according to    
\begin{equation}\label{T0_iter}
T_{0} (1) =  \frac{T_{u}(E_{0}- E_{l}) + T_{l}(E_{u}- E_{0})}{E_{u}-E_{l}} \,.       
\end{equation} 
Using this value in the Gibbs factor, its associated energy $ E_{0}(1)$ should be measured by a MC simulation as the thermal average of the Hamiltonian according to Eq. \eqref{energy_beta_estim}. 

\item 
If $ E_{0}(1) \geq E_{0}$, set $ T_{u} = T_{0} (1) $, $ E_{u} = E_{0}(1)$ and define $ T_m = (T_{l} + T_{0} (1) ) / 2 $, else for $ E_{0}(1) < E_{0}$ set  $ T_{l} = T_{0} (1) $, $ E_{l} = E_{0}(1) $ and define $ T_m = (T_{u} + T_{0} (1) ) / 2 $. Perform a MC simulation at temperature $ T_{m} $ to measure its corresponding energy $ E_{m}$. Now if $ E_{m} \geq E_{0}$ redefine $ T_{u} := T_{m} $ and $ E_{u} = E_{m} $, else for $ E_{m} < E_{0}$ redefine $ T_{l} := T_{m} $ and $ E_{l} := E_{m} $.

Finally, insert the updated values $ T_{l}$, $ E_{l} $ and $ T_{u}$, $ E_{u} $ into the above equation (\ref {T0_iter}) to compute the second approximated value $ T_{0}(2)$, and measure its corresponding energy value $ E_{0}(2) $.

\item Iterate the method going to the step 3 and continue until the energy of the configurations lie within the small thickness spherical shell, which will be achieved when the condition $ ( T_u(n) -T_l(n) ) / T_0(n) \leq \delta $ is fulfilled, for a certain $ \delta $ arbitrary small value.
 
\item When the above inequality is fulfilled, set $n = n_c$ and store the observables. Only those configurations whose energy lies within the small thickness spherical shell centred in $ E_0 $ will actually be included to compute the observables, i.e. configurations $\vec{x}_i$ fulfilling $\mid \mathcal{H}(\vec{x}_i) -E_0\mid / |E_0| \leq \epsilon \, \sigma_E/|\left\langle \right. \hat{E} \left.\right\rangle|$  for $0 < \epsilon < 1$, where $\left\langle \right. \hat{E} \left. \right\rangle$ and $\sigma_E$ are  the average  and standard deviation of the energy estimator. The number of configurations stored should be large enough to obtain trustable statistics, depending on the accuracy required for the computation of the observables.                                                                                                                                                                                                                              
\end{enumerate}

The above explained method allows to find a temperature $T_{0}(n)$ such that $ E_{0}(n) $ lies within a small thickness spherical shell around $ E_{0} $, symbolically $E_{0}(n) \sim E_0 $, and therefore, it generates a sequence of microscopic system configurations with energy arbitrarily close to the input energy $E_0$ by performing within each iteration step $ n $, two conventional  MC simulations in the CE at temperatures $T_{0}(n)$ and $ T_m(n) $ respectively. In other words, it generates a sequence of configurations in an arbitrarily small thickness spherical shell, which corresponds to the right configurations in the MCE for the macroscopic fixed quantities $(E_0,V,N)$. The \textit{third step} of the above explained method ensures, that corresponding temperature values associated to the iterative approximations to $E_{0}(n) \sim E_0 $ also enclose both from below and above the value of $T_0.$ 

It is worthwhile mentioning that obtaining configurations within an arbitrarily small thickness spherical shell in the system phase space, instead of configurations exactly lying on a surface of constant energy $ E_0 $ does not represent a real physical problem, as the thickness can be made as small as required, such that the energy thickness be for instance arbitrary smaller than the energy standard deviation. Moreover, this is exactly the regularization of the Dirac's delta appearing in the distribution probability of the MCE used in standard Statistical Mechanics books \cite{libroGreiner,libroReichl,libroKardar}. This subtle issue arises also in the CE, in which the temperature is held fixed, whose fluctuations nevertheless are non zero, even in high accurate MC simulations. 

The idea underlying the \textit{fourth and fifth steps} relies on the generalized continuity condition of the energy as a function of the temperature at $T = T_0$ , which states that for any arbitrary small value of $ \epsilon $ there is $ \delta $ sufficiently small such that if  $(T_u(n)-T_l(n))/T_0(n) \leq \delta$, it follows that $\mid \mathcal{H}(\vec{x}_i) -E_0\mid / |E_0| \leq \epsilon \, \sigma_E/|\left\langle \right. \hat{E} \left. \right\rangle|$. We use a filtering process based on this condition to ensure that the system configurations used to compute the thermal average of the observables indeed lie within an arbitrarily thin spherical shell around $E_0$.   

In practice and for the efficiency of the method, it is convenient to reach $n_c$ by performing MC simulations with relative small number of sweeps at each $n$-step. In the other hand, since the data stored will be filtered, a rather large statistics for the stored data is required, i.e. sweeps of the order of $10^6$, if high accuracy results are desired. 

It is worthwhile to mention that the histogram technique could be combined with our method. Indeed, the histogram method \cite{FS} and its generalization the multiple histogram technique \cite{FS2}, can be applied in several models and they are especially attractive for the determination of the location and height of peaks of functions. These interesting scanning procedures permit the study of thermodynamic scaling behavior, as well as an accurate determination of corrections to scaling. Nevertheless, since the histogram method requires to generate high accurate histograms of the observables in order to determine the number of configurations for a given set of physical parameters, it cannot be used to calculate observables using improved estimators \cite{libroBinney}, which are generally calculated on the fly, for example   using the so called winding numbers and cluster lengths to calculate the magnetization and magnetic susceptibility  as the cluster (or multicluster) grows, in cluster (multicluster) algorithms in Monte Carlo simulations in quantum spin systems \cite{ELM,BW,PalmaRiveros}. On the other hand, our proposed method allows the implementation of improved estimators of the physical observables and moreover, they can be measured on the fly with high accuracy.

We will illustrate our method by simulating the well-known XY-model defined on a square lattice of lattice size $L$ (see ref. \cite{Estimador} for an accurate analysis of the configurational temperature for this model). The Hamiltonian for this model is: $\mathcal{H} = -JS^2 \sum_{<i,j>}  \cos{(\theta_i-\theta_j)}$, being $J >0$ the ferromagnetic coupling constant, $\theta_i$ is the angle of the spin magnetic moment of magnitude $S$ relative to some direction at the i-th lattice site. The sum is over nearest neighbours and we will use periodic boundary conditions. From now on, the energy will be expressed in units of $JS^2L^2$, i.e. $E = \mathcal{H}/JS^2L^2$. 

\begin{figure}[h!]
  \centering
    \includegraphics[width=\linewidth]{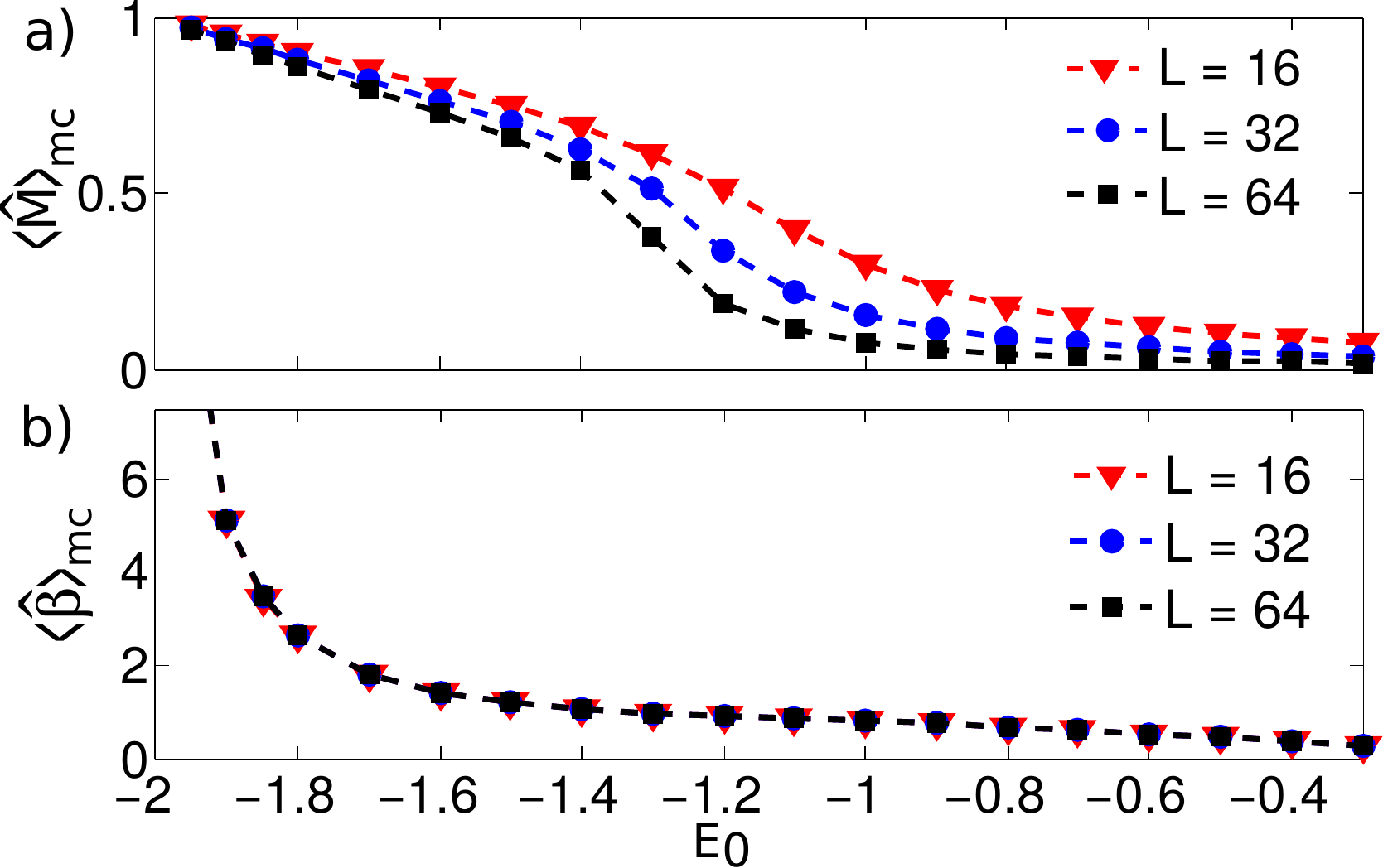}
  \caption{\textit{(color online) Average of the a) magnetization and b) inverse temperature estimators as a function of the energy of the 2d XY-model in a square lattice of lattice sizes $L = 16$, $32$ and $64$, for configurations within the spherical shell around $ E_0 $. }}
  \label{observables_fig}
\end{figure}

For each value $T_0(n)$  calculated by using Eq.\eqref{T0_iter} and $T_m(n)$, ($n = 1,2, \cdots$) we have performed a MC runs in a square lattice of lattice size $L = 32$, in the CE using the Metropolis algorithm with over-relaxation, performing 1000 sweeps until the condition of the \textit{fourth step} is fulfilled (at $n = n_c$), using $\delta = 0.00001$. Then a MC simulation at $T_0(n_c+1)$ is performed using a large number of sweeps, $\mathcal{N} = 2 \times 10^6$, to ensure a large enough statistical data,  storing the data for the observables, and finally, we have set $\epsilon = 0.5$ to define the thickness of the spherical shell. The over-relaxation were performed each 100 sweeps and in order to estimate the finite-size effects, in adittion we have used the lattice sizes $ L = 16 $ and 64 (see figure \ref{observables_fig}).

\begin{figure}[H]
  \centering
    \includegraphics[width=\linewidth]{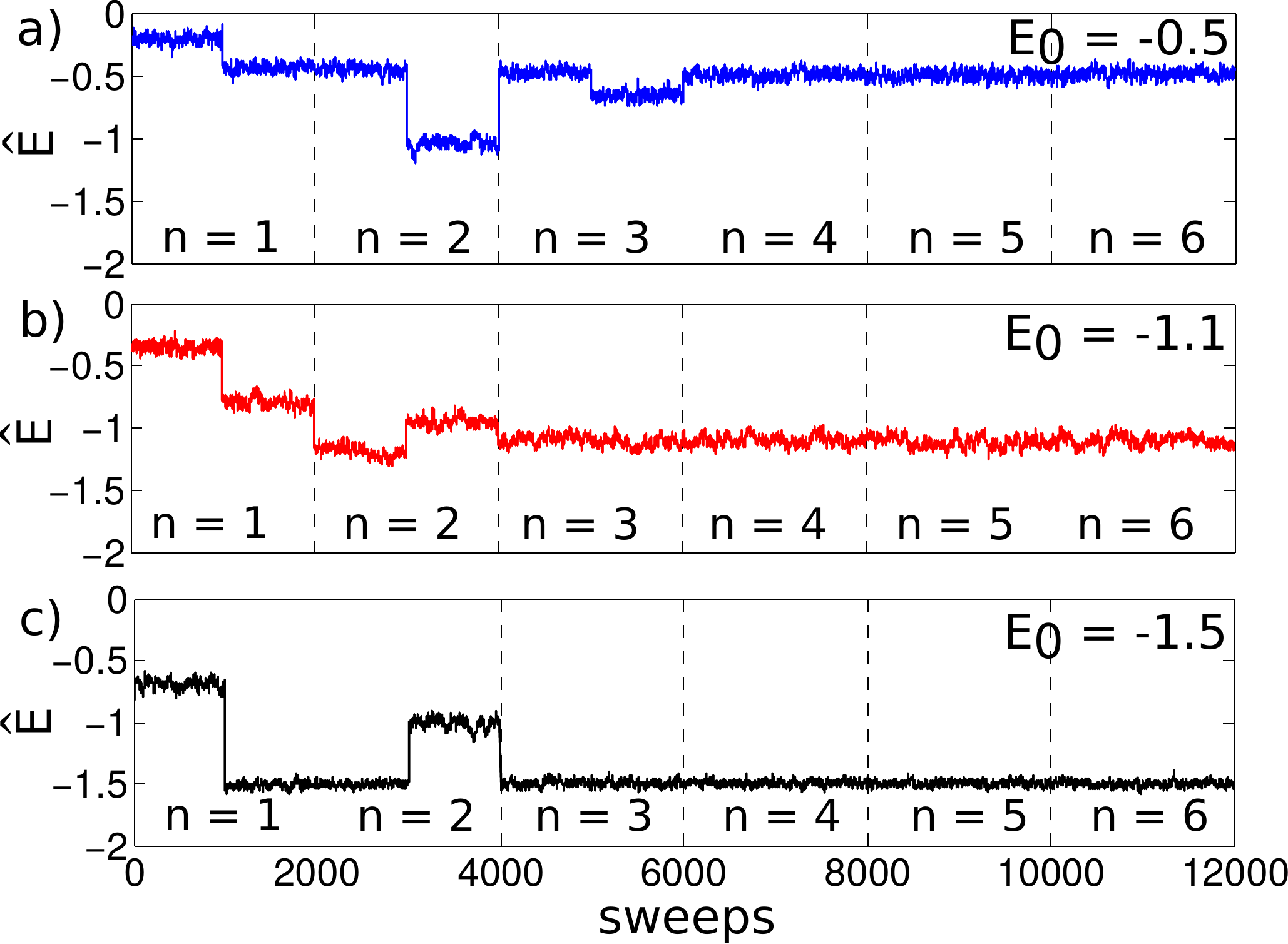}
  \caption{\textit{(color online) Relaxation path of the energy as the $n$-step increases, displayed for different input energies a) $E_0 = -0.5$, b) $E_0 = -1.1$ and c) $E_0 = -1.5$. The XY-model defined on a square lattice of size $L = 32$ was simulated. As it can be seen, the method converges to $E_0$ as $n$-increases. }}
  \label{interpolation_converge}
\end{figure}

Fig.\ref{interpolation_converge} shows the thermalization path of the energy estimator as the $n$-iteration number of the method increases for three given energy inputs a) $E_0 = -0.5$, b) $E_0 = -1.1$, and c) $E_0 = -1.5$, and for initial low and upper temperature bounds $T_l = 0.01$ and $T_u = 6.0$ respectively. It can be seen that as the number of interpolations increases, the thermalization path converges faster to $E_0$, which happens even for a relative wide range of the bounds $[T_l,T_u]$.

\begin{figure}[h!]
  \centering
    \includegraphics[width=\linewidth]{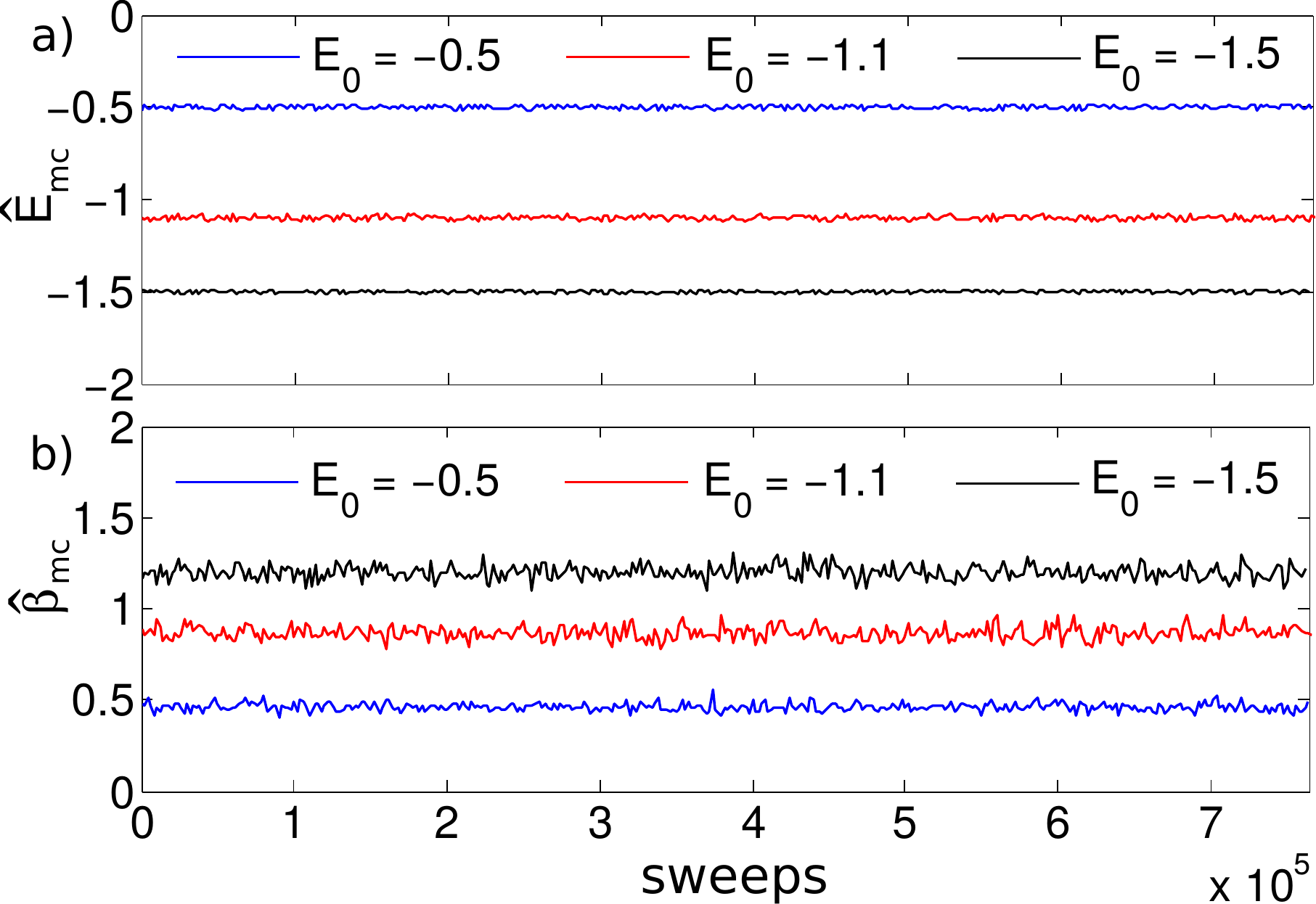}
  \caption{\textit{(color online) Time series of the  a) energy  and b) inverse temperature estimators of the configurations within the spherical shell thickness fulfilling $\mid \mathcal{H}(\vec{x}_i) -E_0\mid / |E_0| \leq \, \sigma_E/2 |\left\langle \right. \hat{E} \left. \right\rangle|$ for the XY-model in a square lattice size $L = 32$ for different input energies $E_0$.}}
  \label{fluctuation_energy_beta}
\end{figure}

In Fig. \ref{fluctuation_energy_beta} the time series of the  a) energy and b) inverse temperature estimators in the MCE are displayed, i.e. only those microscopic configurations lying within the thin spherical shell around $E_0$ were included. This was achieved by filtering out the $\mathcal{N}$ data using $\epsilon = 0.5$ as explained above in the \textit{fifth step}. Different values for $E_0$ were used to show the robustness of the convergence even in the low-temperature region, where the XY-model has an infinite correlation length in the thermodynamic limit.

The consistency of the method can be directly checked by measuring the average of the energy, which should be $E_0$ up to very small deviations corresponding to the thickness used for the spherical shell defined in \textit{fifth step}. Table I displays precisely the average of the Hamiltonian (energy of the system) for different input energies $E_0$, which allows to conclude the high accuracy measured for the thermal average values, as the deviation respect to the input values are of the order of a few per ten thousand  (as shown in the third column by the relative error). The last column shows the acceptance percent of the data within the spherical shell respect to the whole data of length $\mathcal{N}$. The accuracy can be further improved by increasing the configuration numbers and/or decreasing the thickness of the spherical shell if required.
\begin{table}[h!]
\label{numerical_details}
\centering
\begin{tabular}{|c|ccc|}

\multicolumn{4}{l}{}   \\  \hline
 
$E_0$        &$\langle \hat{E} \rangle_\text{mc}$      & rel. error   & acc$\%$  \\ \hline

-0.3000 & -0.2999 &  0.0003    & 38.33 \\

-0.5000 & -0.5001 &  0.0002    & 38.21 \\

-0.7000 & -0.6999  & 0.0001   &  38.31    \\  

-1.1000 & -1.0997 & 0.0003  & 38.32    \\

-1.5000 & -1.5002 & 0.0001  & 38.18 \\

-1.7000 & -1.7001 & 0.0001  & 38.28 \\ \hline

\end{tabular}
\caption{Average of the energy estimator for the sampled configurations compatible with the energy $E_0$ (first column) and its corresponding relative error (third column) for a lattice size $L = 32$. The last column shows the percentage of configurations fulfilling $\mid \mathcal{H}(\vec{x}_i) -E_0\mid / |E_0| \leq \sigma_E/2 | \left\langle \right. \hat{E} \left. \right\rangle|$.} 
\end{table}

In Fig. \ref{observables_fig} the average of the a) magnetization and b) inverse of temperature estimators are shown as a function of the input energy $ E_0 $ for three different lattice sizes, $ L =$ 16, 32 and 64. The thermal averages were computed using only those configurations $\vec{x}_i$, which fulfil the condition of the \textit{fifth step}: $\mid \mathcal{H}(\vec{x}_i) -E_0\mid / |E_0| \leq \sigma_E/2 |\left\langle \right. \hat{E} \left. \right\rangle|$. The upper curve shows the finite-size effects of the magnetization, while the lower one shows the inverse of the temperature $\beta$. Note that for each input energy value $E_0$, the corresponding $\beta$-values for the different system sizes collapse onto the same point, which is consistent with the fact that the temperature represents an intensive quantity.

It should be mentioned that a larger statistics for the observables than the computed by using configurations whose energy lies within the thin spherical shell centered at $E_0$ could be obtained, by performing several Microcanonical MC simulations different input energies $E_i$, and handle all the data using the multiple histogram method. The collected data would give physical information of the observables in a region (for instance, the scaling region, close to a critical point). On the contrary, if one is interested to measure some observables at some fixed energy value $E_0$, the final filtering procedure is necessary to obtain results within the required numerical accuracy, as explain in the \textit{fifth-step} of our method.

As a summary, we have presented a general Microcanonical ensemble MC method, which relies at each iteration step $E_{0}(n) \rightarrow E_0 $ on MC simulations in the CE, and therefore inherits the ergodicity and microscopic reversibility  properties underlying the algorithm used. Moreover it has, as a consequence, the advantage of allowing the use of efficient algorithms developed for canonical ensemble MC simulations, for instance to accurately study critical systems close to transition points. 

The method relies on the existence of a locally one-to-one equation of state $ E = E(T) $, and generates a sequence of configurations of the system whose energy lies within an arbitrarily small thickness spherical shell centered in $ E_0 $, and does not require any extra physical degree of freedom, and its structure allows to be implemented for arbitrary physical models.

We have illustrated its efficiency by simulating the two-dimensional XY-model in a wide range of temperature values, including the low T-region below the Berezinskii-Kosterlitz-Thouless critical temperature, which is known to have a continuous line of critical points with infinite correlation length in the thermodynamic limit, and thus local-update based numerical simulations will suffer from the so-called critical slowing down phenomenon \cite{libroBinney}. We want to stress that MC simulations in the MCE could be used to address the challenging issue of the ergodic dynamics and thermalizations in quantum isolated systems \cite{Neill}, as the energy in these systems is a conserved quantity.

It is worthwhile mentioning that an extension of the current method to connect microcanonical MC simulations to simulations in other ensembles should be straightforward by having the required configurational estimators for the conjugated macroscopic variables, and following the lines discussed in the description of the proposed method.

This work was partially supported by Dicyt-USACH Grant No. 041831PA. A.R. acknowledges support from Fondecyt under Grant 3180470.

\bibliographystyle{ieeetr}
\bibliography{referencias}

\end{document}